\documentclass[showpacs,a4paper,fleqn]{revtex4-1}
\usepackage{graphicx}
\usepackage[fleqn]{amsmath}
\usepackage{amsmath}
\def\bm#1{\hbox{\boldmath$#1$\unboldmath}}

\def\rmi{{\rm{i}}}

\begin{document}

\title{General spin and pseudospin symmetries of the Dirac equation}

\author{P. Alberto}
\affiliation{CFisUC, Physics Department, University of Coimbra,
P-3004-516 Coimbra, Portugal}
\author{M. Malheiro, T. Frederico}
\affiliation{Instituto Tecnol\'ogico de Aeron\'autica, DCTA,
12228-900 S\~ao Jos\'e dos Campos, S\~ao Paulo, Brazil}
\author{A. de Castro}
\affiliation{Departamento de F{\'\i}sica e Qu{\'\i}mica,
Universidade Estadual Paulista, 12516-410 Guaratinguet\'a, S\~ao
Paulo, Brazil}
\date{\today}

\begin{abstract}
In the 70's Smith and Tassie, and Bell and Ruegg independently found SU(2) symmetries of the Dirac equation with scalar and vector potentials. These symmetries, known as pseudospin and spin symmetries, have been extensively researched and applied to several physical systems. Twenty years after, in 1997, the pseudospin symmetry has been revealed by Ginocchio as a relativistic symmetry of the atomic nuclei when it is described by relativistic mean field hadronic models.
The main feature of these symmetries is the suppression of the spin-orbit coupling either in the upper or lower components of the Dirac spinor, thereby turning the respective second-order equations into Schr\"odinger-like equations, i.e, without a matrix structure.
In this paper we propose a generalization of these SU(2) symmetries for potentials in the Dirac equation with
several Lorentz structures, which also allow for the suppression of the matrix structure of
second-order equation equation of either the upper or lower components of the Dirac spinor. We derive the general
properties of those potentials and list some possible candidates, which include the usual spin-pseudospin
potentials, and also 2- and 1-dimensional potentials.
An application for a particular physical system in two dimensions, electrons in graphene, is suggested.
\end{abstract}

\maketitle

\section{Introduction}

Pseudospin symmetry has been a topic in nuclear physics since the late 60's,
when it was introduced to explain the near degeneracy of some single-particle
levels near the Fermi surface.
The subject was revived in 1997 when Ginocchio was able to relate it
with a symmetry of the Dirac equation with scalar $S$ and vector $V$ mean-field
potentials such that $V=-S+ C$ where $C$ is a constant \cite{gino_1997}.
A related symmetry, the spin
symmetry, was used to explain the suppression of spin-orbit splittings
in states of mesons with a heavy and a light quark. Actually, both symmetries had been
 found before in the 70's independently by Smith and Tassie \cite{smith} and by Bell and Ruegg \cite{bell} as SU(2) symmetries
 of the Dirac equation with scalar and vector potentials, i.e., potentials coupling to the mass and energy, respectively.
Two reviews of the subject, one by Ginocchio \cite{gino_rev_2005} and a recent one by Liang \textit{et al.} \cite{Liang_Meng_Zhou_rev},
give a good overall
account of the many results produced in these last 18 years, regarding both the main features
of those symmetries and their applications to physical systems.

One notable feature of these symmetries is the suppression of either the spin-orbit or
the so-called pseudospin-orbit coupling that are present in the second-order equations
for the upper and lower Dirac spinor components, respectively. Since those terms arise from
the coupling of those spinor components in their first-order Dirac equations,
they have a non-trivial, i.e., different from identity,
matrix structure. Therefore, their suppression amounts to have the upper (spin symmetry) or lower
(pseudospin symmetry) spinors satisfying second-order equations of Shr\"odinger type, i.e, with no
 matrix structure
(see  \cite{prl_86_5015,harm_osc_prc_2004_2006, tensor,spectra,pedro_pra}, and \cite{conf_Praga} for a brief review).
This is possible because we have both scalar and vector potentials $S$ and $V$. In particular, when one is close to
spin symmetry conditions ($S\sim V$), one can suppress spin-orbit couplings even in a relativistic fermion system.
This may come as a surprise at first, since spin-orbit coupling can be shown to be
a relativistic correction to non-relativistic quantum mechanics with only vector potentials \cite{palberto_LS}.
It is worthwhile to remark that these symmetries give also rise to supersymmetric patterns
of the Dirac Hamiltonian \cite{Leviatan_prl}.

In this paper we aim to investigate which are the conditions for a pair of potentials in Dirac equation
to produce this same effect, i.e., the suppression of the matrix structure for second-order
equations for either the upper or the lower component of the Dirac spinor.
Thus, we are generalizing the findings of Smith and Tassie \cite{smith}, and  Bell and Ruegg \cite{bell},
where only the case of a pair of scalar and vector potentials was considered.

These conditions are derived in Section \ref{section 2}, as well as the generators of the corresponding SU(2)
symmetries, showing at the same time that they commute with the Dirac Hamiltonian.
In the following Section we discuss the Lorentz structure of potentials satisfying those conditions and weaker
conditions as well, leading in this last case to 2- and 1-dimensional potentials in coordinate space.

Finally we draw the conclusions and briefly discuss possible physical systems to which the previous results may be applied,
and in particular the case of electrons in graphene, a two-dimensional system described by the Dirac equation of
relativistic fermions with effective mass zero, which in principle could also exhibit such symmetries.

\section{General Spin and pseudospin symmetries in the Dirac equation}
\label{section 2}
\subsection{Generators of the symmetry}

The time--independent Dirac equation for a spin 1/2 particle with energy $E$, under
the action of an external hermitian $V$ potential (which may include a mass term) with a general Lorentz structure reads
\begin{equation}\label{eq:dirac_V}
H\psi=(\bm\alpha\cdot \hat{\bm p} + V)\psi=E\psi\ .
\end{equation}
where $\alpha_i$ ($i=1,2,3$) are the $4\times 4$ matrices related to the usual Dirac gamma matrices $\gamma^\mu$, $\mu=0,1,2,3$,  by $\alpha_i=\gamma^0\gamma^i$. Units $\hbar=c=1$ are used.

We will now consider the conditions for the potential $V$ under which the Dirac equation (\ref{eq:dirac_V})
has  SU(2) symmetries of the type described by Bell and Ruegg \cite{bell}, following closely their procedure.

Let us consider operators $P_\pm=(I\pm O)/2$ in the spinor 4-dimensional space
which have the projector properties
\begin{eqnarray}
\label{proj_prop1}
P_\pm P_\pm&=&P_\pm  \\
\label{proj_prop2}
P_\pm P_\mp&=&0 \ ,
\end{eqnarray}
which are satisfied if the matrix $O$ is such that $O^2=I$, $I$ being the identity matrix
in spinor space. We further require that
\begin{equation}\label{anticomm1}
\{\alpha_i,O\}=0 \ ,
\end{equation}
which implies that
\begin{equation}\label{anticomm2}
P_\pm\alpha_i=\alpha_i P_\mp \ .
\end{equation}

If $V$ has the general form $V=V_O O +V_v I$, where $V_O$ and $V_v$ are functions of the coordinates,
it can be written as
\begin{equation}\label{V}
    V=V_+P_++V_-P_-\qquad{\rm with}\qquad V_\pm=V_v\pm V_O \ .
\end{equation}

We apply now the projectors $P_\pm$ to the Dirac equation (\ref{eq:dirac_V}) to get the two coupled equations
\begin{eqnarray}
\label{psi-}
 \bm\alpha\cdot\hat{\bm p}\,\psi_-+V_+\,\psi_+&=&E\psi_+ \\
\label{psi+}
 \bm\alpha\cdot\hat{\bm p}\,\psi_++V_-\,\psi_-&=&E\psi_- \ ,
\end{eqnarray}
where $\psi_\pm=P_\pm\psi$.

Let us now set one of the potentials $V_\pm$, say $V_-$, to a constant $C_-$, meaning that $V_v=V_O+C_-$.
Applying $\bm\alpha\cdot\hat{\bm p}$ to equation (\ref{psi+}), and using the general property
\begin{equation}\label{alpha_rel}
 \bm\alpha\cdot{\bm A}\,\bm\alpha\cdot{\bm B}={\bm A}\cdot{\bm B}+\rmi({\bm A}\times{\bm B})\cdot{\bm\Sigma}
\end{equation}
where $\bm\Sigma=\bm\alpha\times\bm\alpha/(2\rmi)=\gamma^5\bm\alpha$ is the spin matrix in four-dimensional spinor space,
we get
\begin{equation}\label{psi+2}
\hat{\bm p}^2\,\psi_+=(E-C_-)\bm\alpha\cdot\hat{\bm p}\,\psi_-=(E-C_-)(E-V_+)\,\psi_+
\end{equation}
using also eq. (\ref{psi-}).
This is a Schr\"odinger-type equation for $\psi_+$ with no matrix structure. Therefore, it is invariant under
the infinitesimal spin transformation
\begin{equation}\label{delta_psi+}
\psi_+\to \psi_++\delta\psi_+=\psi_++\frac{\bm\epsilon\cdot\bm\Sigma}{2\rmi}\psi_+
\end{equation}
The corresponding transformation for $\psi_-$, using (\ref{psi+}) and (\ref{alpha_rel}), is
\begin{eqnarray}
\nonumber
  \delta\psi_-&=&\frac{\bm\alpha\cdot\hat{\bm p}}{E-C_-}\,\delta\psi_+
  =\frac{\bm\alpha\cdot\hat{\bm p}}{E-C_-}\,\frac{\bm\epsilon\cdot\bm\Sigma}{2\rmi}\psi_+ \\
  \nonumber
   &=&\bm\alpha\cdot\hat{\bm p}\,\frac{\bm\epsilon\cdot\bm\Sigma}{2\rmi}
   \frac{\bm\alpha\cdot\hat{\bm p}\,\,\bm\alpha\cdot\hat{\bm p}}{\hat{\bm p}^2\,(E-C_-)}\,\psi_+  \\
   &=& \frac{\bm\epsilon}{2\rmi}\cdot \frac{\bm\alpha\cdot\hat{\bm p}\,\bm\Sigma\,\bm\alpha\cdot\hat{\bm p}}
   {\hat{\bm p}^2}\,\psi_-\  .
\end{eqnarray}
For the transformation of the full spinor $\psi=\psi_++\psi_-$, and defining
$\bm s={\bm\alpha\cdot\hat{\bm p}\,\bm\Sigma\,\bm\alpha\cdot\hat{\bm p}}\big/{\hat{\bm p}^2}$, we get
\begin{equation}\label{delta_psi}
    \delta\psi=\delta\psi_++\delta\psi_-=
    \frac{\bm\epsilon}{2\rmi}\cdot(\bm\Sigma\psi_++\bm s\psi_-)=
    \frac{\bm\epsilon}{2\rmi}\cdot(\bm\Sigma\, P_++\bm s\,P_-)\psi \ ,
\end{equation}
from which we can write the generator of this transformation as
\begin{equation}\label{gen_sym-}
\bm S_-=\bm\Sigma\, P_++\bm s\,P_- \ .
\end{equation}

One can obtain the second-order equation for $\psi_-$ from eqs. (\ref{psi-}) and (\ref{psi+}), using (\ref{alpha_rel}) again.
It reads
\begin{equation}\label{psi-2}
\hat{\bm p}^2\,\psi_-+\frac{1}{E-V_+}\big(\nabla V_+\times\hat{\bm p}\cdot{\bm\Sigma}-\rmi\,\nabla V_+\cdot\hat{\bm p}\big)\psi_-=(E-C_-)(E-V_+)\,\psi_- \ .
\end{equation}
If the potential $V_+$ is radial, in the second term in the left-hand side of the equation we can identify
 a spin-orbit coupling term and the Darwin term \cite{conf_Praga}.

\subsection{Commutation with the Dirac Hamiltonian}

One can check that the generator $\bm S_-$ of the transformation described above is indeed a symmetry operator
by computing its commutator with the Hamiltonian
\begin{equation}\label{H-}
    H_-=\bm\alpha\cdot\hat{\bm p}+V_+P_++C_-P_-  \ .
\end{equation}
One may note that condition (\ref{anticomm1}) implies that
\begin{equation}\label{cond_3}
    [O,\bm\Sigma]=[P_\pm,\bm\Sigma]=0 \ ,
\end{equation}
provided that $\big\{\gamma^5,O\big\}=0$, which is true as long as $O$ contains an odd number of distinct $\gamma^\mu$ matrices.
Actually, this last requirement is also a necessary condition for (\ref{anticomm1}) to hold.

One has
\begin{equation}\label{commut_geral}
[H_-,(S_-)_i]=[\bm\alpha\cdot\hat{\bm p},(S_-)_i]+[V_+P_+,(S_-)_i]+[C_-P_-,(S_-)_i]\,,\quad i=1,2,3.
\end{equation}
For the last term one has
\begin{equation}\label{comm1}
    [C_-P_-,(S_-)_i]=C_-\,[P_-,(S_-)_i]=C_-\,[P_-,\Sigma_i\,P_+]+C_-\,[P_-,s_i\,P_-]=0 \  ,
\end{equation}
since $P_-$ commutes with $\Sigma_i$ and $s_i$ and also because of property (\ref{proj_prop2}).

For the second term in (\ref{commut_geral})
\begin{equation}\label{comm2}
    [V_+P_+,(S_-)_i]=V_+\,[P_+,(S_-)_i]+[V_+,(S_-)_i]P_+=0 \  .
\end{equation}
Here the first term is zero because $P_+$ commutes with $\Sigma_i\,P_+$ and $s_i\,P_-$.
The second term is also zero because, (1) $\Sigma_i$ commutes with $V_+$, (2) the $s_i$ term contains
the product $P_-P_+$, so $[V_+P_+,(S_-)_i]=0$.

Finally, for the first term in (\ref{commut_geral}) we have
\begin{equation}\label{comm3}
    [\bm\alpha\cdot\hat{\bm p},(S_-)_i]=[\bm\alpha\cdot\hat{\bm p},\Sigma_i\,P_+]+
    [\bm\alpha\cdot\hat{\bm p},s_i\,P_-]  \  .
\end{equation}
Furthermore, we have
\begin{eqnarray}
\label{comm31}
\nonumber
    [\bm\alpha\cdot\hat{\bm p},\Sigma_i\,P_+]&=&[\bm\alpha\cdot\hat{\bm p},\Sigma_i]P_++
    \Sigma_i\,[\bm\alpha\cdot\hat{\bm p},P_+]\\
    &=&[\bm\alpha\cdot\hat{\bm p},\Sigma_i]P_++
    \Sigma_i\,\bm\alpha\cdot\hat{\bm p}(P_+-P_-)=
        [\bm\alpha\cdot\hat{\bm p},\Sigma_i]P_++
    \Sigma_i\,\bm\alpha\cdot\hat{\bm p}\,O \ .
\end{eqnarray}
Similarly,
\begin{eqnarray}\label{comm32}
\nonumber   [\bm\alpha\cdot\hat{\bm p},s_i\,P_-]&=&[\bm\alpha\cdot\hat{\bm p},s_i]P_-+
    s_i\,[\bm\alpha\cdot\hat{\bm p},P_-]=  [\bm\alpha\cdot\hat{\bm p},s_i]P_-+
    s_i\,\bm\alpha\cdot\hat{\bm p}(P_--P_+)\\
  &=&        [\bm\alpha\cdot\hat{\bm p},s_i]P_--
    s_i\,\bm\alpha\cdot\hat{\bm p}\,O = [\Sigma_i,\bm\alpha\cdot\hat{\bm p}]P_--
    \bm\alpha\cdot\hat{\bm p}\,\Sigma_i\,O  \ .
\end{eqnarray}
Summing these two last expressions we get
\begin{equation}\label{comm33}
      [\bm\alpha\cdot\hat{\bm p},(S_-)_i] = [\bm\alpha\cdot\hat{\bm p},\Sigma_i](P_+-P_-)+
       [\Sigma_i,\bm\alpha\cdot\hat{\bm p}]\,O=[\bm\alpha\cdot\hat{\bm p},\Sigma_i]\,O+
       [\Sigma_i,\bm\alpha\cdot\hat{\bm p}]\,O=0 \ .
\end{equation}
This completes the proof that $[H_-,(S_-)_i]=0$.

\subsection{Algebra of the generators. The spin symmetry case}

Let us consider the commutation relations between each component of generator $\bm S_-$, namely,
\begin{equation}\label{commut1}
  [(S_-)_i,(S_-)_j]\qquad i,j=1,2,3\quad .
\end{equation}
One has
\begin{equation}\label{commut2}
  [(S_-)_i,(S_-)_j]=[\Sigma_i\, P_++s_i\,P_-,\Sigma_j\, P_++s_j\,P_-]=[\Sigma_i,\Sigma_j]\, P_++[s_i,s_j]\,P_- ,
\end{equation}
because of the projector properties (\ref{proj_prop1}) and (\ref{proj_prop2}) and because $P_\pm$ commutes with $\Sigma_i$ and $s_i$.
Since
\begin{equation}\label{commut3}
  [\Sigma_i,\Sigma_j]=2\rmi\,\varepsilon_{ijk}\Sigma_k\qquad{\rm and}\qquad [s_i,s_j]={\bm\alpha\cdot\hat{\bm p}\,[\Sigma_i,\Sigma_j]\,\bm\alpha\cdot\hat{\bm p}}\big/{\hat{\bm p}^2}=2\rmi\,\varepsilon_{ijk}s_k \ ,
  \end{equation}
one has
\begin{equation}\label{commut4}
  [(S_-)_i,(S_-)_j]=2\rmi\,\varepsilon_{ijk}(S_-)_k
   \end{equation}
i. e., the generators $(S_-)_i$ satisfy a SU(2) algebra. This case,  $V_v=V_O+C_-$, is usually known as the spin symmetry case of the Dirac Hamiltonian, since we have a normal spin transformation in the upper component of the Dirac spinor as shown in \eqref{delta_psi+}.

\subsection{The other symmetry: the pseudospin symmetry case}

Of course, we could as well have set instead $V_+$ to a constant $C_+$. In that case, the roles of $\psi_\pm$  would
be reversed and one would have another symmetry whose generator would be
\begin{equation}\label{gen_sym+}
\bm S_+=\bm\Sigma\, P_-+\bm s\,P_+ \ ,
\end{equation}
which would commute with the Hamiltonian
\begin{equation}\label{H+}
    H_+=\bm\alpha\cdot\hat{\bm p}+V_-P_-+C_+P_+ \  .
\end{equation}
Similarly as was done in the previous subsection, one can show that these generators also satisfy a SU(2) algebra.

The second-order equations for the upper and lower spinors would then be
\begin{eqnarray*}
\hat{\bm p}^2\,\psi_+&+&\frac{1}{E-V_-}\big(\nabla V_-\times\hat{\bm p}\cdot{\bm\Sigma}-\rmi\,\nabla V_-\cdot\hat{\bm p}\big)\psi_+=(E-C_+)(E-V_-)\,\psi_+ \\
\hat{\bm p}^2\,\psi_-&=&(E-C_+)(E-V_-)\,\psi_- \ .
\end{eqnarray*}

This case,  $V_v=-V_O+C_+$, is usually known as the pseudospin symmetry case of the Dirac Hamiltonian, since we have a normal spin transformation in the lower component of the Dirac spinor which has an inverse parity relative to the one of the upper component and to the parity of the whole spinor.

\section{Possible $O$ matrices}

From the previous section, in order to have one of these two SU(2) symmetries, the matrix $O$ must satisfy the following relations:
\begin{enumerate}
  \item $O^2=I$
  \item $\{\alpha_i,O\}=0$ \ .
  \item $[O,\Sigma_i]=0$ \ .
\end{enumerate}
As explained before, the relation $[O,\Sigma_i]=0$ is a consequence of condition 2.
These three requirements are satisfied by the Hermitian matrices $\beta=\gamma^0$ and $\rmi\gamma^0\gamma^5$.
The case of $O=\gamma^0$ leads to the well-known spin and pseudospin symmetries described in the Introduction.

If one relaxes the second requirement, one can also consider the linear combination $O=\bm\lambda\cdot\bm\alpha$,
such that $\bm\lambda$ is a constant unit vector ($\bm\lambda\cdot \bm\lambda=1$) and also such that
$\bm\lambda\cdot \hat{\bm p}\,\,\psi=0$. Because one has $P_\pm\bm\alpha\cdot \hat{\bm p}\,\,\psi=\bm\alpha\cdot \hat{\bm p}P_\mp\,\psi
\pm \bm\lambda\cdot \hat{\bm p}\,\,\psi$, one would still have $P_\pm\bm\alpha\cdot \hat{\bm p}\,\psi=\bm\alpha\cdot \hat{\bm p}P_\mp\,\psi$.
Then the third condition can be satisfied in a weak way, considering
transformations with infinitesimal parameters $\bm\epsilon$ such that
\begin{equation}\label{3req}
    [\bm\lambda\cdot\bm\alpha,\bm\epsilon\cdot\bm\Sigma]=2\rmi\,(\bm\lambda\times\bm\epsilon)\cdot\bm\alpha=0\ ,
\end{equation}
meaning that the vectors $\bm\lambda$ and $\bm\epsilon$ must be parallel. For instance, if $\bm\lambda=\hat e_z$,
i.e., $O\equiv\alpha_3=\gamma^0\gamma^3$, then one should have $\hat{p}_3\,\psi=0$ and $\bm\epsilon=\epsilon \hat e_z$. In this
case the symmetry generator would be the matrix $\Sigma_3$, generator of the two-dimensional rotation group in four-component
spinor space, which is a realization of the unitary group $U(1)$. Our problem would be 2-dimensional, i.e.,
the spinor (and potentials), would depend only on coordinates $x,\,y$.

Another possibility for $O$ would be the linear combination of the space components of the tensor operator in spinor space
$\gamma^0\sigma^{0i}=\rmi\beta\alpha_i$, i.e., $O=\rmi\beta\bm\lambda\cdot\bm\alpha$. The first requirement would be met
again by setting $\bm\lambda\cdot\bm\lambda=1$. The second and third requirements would be met by setting, respectively,
\begin{flalign}\label{2req2}
  &\{\beta\bm\lambda\cdot\bm\alpha,\bm\alpha\cdot\hat{\bm p}\}\,\psi=
    \beta[\bm\lambda\cdot\bm\alpha,\bm\alpha\cdot\hat{\bm p}]\,\psi
    =2\rmi\,\beta\bm\lambda\times\hat{\bm p}\cdot\bm\Sigma\,\psi=0\ ,\\[2mm]
\label{3req2}
    &[\beta\bm\lambda\cdot\bm\alpha,\bm\epsilon\cdot\bm\Sigma]=\beta[\bm\lambda\cdot\bm\alpha,\bm\epsilon\cdot\bm\Sigma]
    =2\rmi\,\beta\bm\lambda\times\bm\epsilon\cdot\bm\alpha=0 \ .
\end{flalign}

The first condition would be satisfied if $\bm\lambda\times\hat{\bm p}\,\psi=0$ and the second one if $\bm\epsilon$ is
parallel to $\bm\lambda$. If one chooses again $\bm\lambda=\hat e_z$, this would give rise to a 1-dimensional potential,
depending only on $z$.

\section{Discussion and conclusions}

We have derived the general conditions under which a general potential plus a vector potential give rise to spin and pseudospin-like
symmetries in the Dirac equation, i.e., lead to a Schr\"odinger-like equation for the upper or lower component of the Dirac spinor.
In three-dimensional space, we showed that there are two potentials which satisfy those conditions: a scalar potential, giving rise to the usual the spin and pseudospin symmetries found independently by Smith and Tassie \cite{smith}, and  Bell and Ruegg \cite{bell} , and a pseudoscalar potential. In this last case the Dirac Hamiltonian would read
\begin{equation}\label{Hamilt_dirac_gamma5_V}
H=\bm\alpha\cdot \hat{\bm p} + \rmi\beta\gamma^5 V_{ps}+ V_v\ ,
\end{equation}
with $V_{ps}=\pm V_v$. In physical terms, this would correspond to a system of massless fermions interacting with
mean-field pseudoscalar and vector potentials which have the same magnitude. One physical system in which this symmetry would be slightly broken would describe a fermion, say, a baryon, with a relatively small (effective) mass, interacting with a pion and $\omega$ meson.

As was shown in the last section, if we constrain the fermion motion to 2- and 1-dimensional space, there are additional
potentials for which these symmetries can be realized. The respective spin or pseudospin symmetric
Hamiltonians would look like
\begin{eqnarray}
\label{Hamilt_2}
  H_2&=&\alpha_x\hat p_x +\alpha_y\hat p_y+ \alpha_z V_z+ V_{2v} \\
\label{Hamilt_1}
  H_1&=&\alpha_z\hat p_z + \rmi\beta\alpha_z V_{t}+ V_{1v} \ ,
\end{eqnarray}
with 2- and 1- dimensional mean-field potentials such that $V_z(x,y)=\pm V_{2v}(x,y)$ and $V_t(z)=\pm V_{1v}(z)$.

Equation (\ref{Hamilt_2}) can represent a particularly interesting way of realizing these symmetries in two-dimensional coordinate space,
 because there is indeed a physical system with relativistic fermions with effective mass zero which could exhibit such symmetries: electrons in graphene. These effective particles can be described by a massless 3+1 Dirac equation within the framework of interacting quantum field-theories (see e.g. \cite{JacPRL07,OliPRB11}). One example is the continuum spectrum of the Dirac electron interacting with two dimensional potentials embedded in a 3+1 space \cite{grafeno}. Again, in that theory one has in general also potentials with Lorentz structure other than
vector, and, in this case, the third component of a four-vector potential $(\gamma^0\gamma^3=\alpha_z)$
(note that the Lorentz character of $V_O O$ is given by its form in the  covariant form of the Dirac equation, i.e., $\gamma^0 V_O O$),
 but it may be interesting to assess the effect of this symmetry on the continuum and discrete spectrum of the Dirac electrons as well as its breaking due the other potentials. In this way, the Dirac electrons on graphene could be a tool to study the consequences of the  generalized spin and pseudospin symmetries in a controllable form. We leave for a future work such detailed investigation.

\begin{acknowledgments}
PA would like to thank the Universidade Estadual
Paulista, Guaratinguet\'a Campus, for supporting his stays in its
Physics and Chemistry Department. MM and TF would like to thank FAPESP for support under the thematic project 2015/26258-4 and CNPq for partial support.

\end{acknowledgments}

\end{document}